\documentclass[twocolumn,citeautoscript,prl,superscriptaddress,amsmath,amssymb,8pt]{revtex4-1}
\usepackage{graphicx}
\usepackage{dcolumn}
\usepackage{bm}% bold math

\usepackage[usenames,dvipsnames]{xcolor} \usepackage{dsfont}
 \usepackage{amsbsy}

\usepackage{graphicx,bm,times}
 \usepackage{amsmath,amssymb} 
\usepackage{dsfont}
\usepackage{appendix}
\date{\today}

\begin{document}

\title{Adiabatic Dynamical-Decoupling Based Control of Nuclear Spin Registers }
\author{O.T. Whaites} \affiliation{Department of Physics and Astronomy, University College London,
Gower Street, London WC1E 6BT, United Kingdom}
\author{J.  Randall } \affiliation{QuTech, Delft University of Technology, PO Box 5046, 2600 GA Delft, The Netherlands}
\affiliation{Kavli Institute of Nanoscience Delft, Delft University of Technology, PO Box 5046, 2600 GA Delft, The Netherlands}
\author{T.H. Taminiau } \affiliation{QuTech, Delft University of Technology, PO Box 5046, 2600 GA Delft, The Netherlands}
\affiliation{Kavli Institute of Nanoscience Delft, Delft University of Technology, PO Box 5046, 2600 GA Delft, The Netherlands}
\author{T.S. Monteiro} \affiliation{Department of Physics and Astronomy, University College London,
Gower Street, London WC1E 6BT, United Kingdom}

\begin{abstract} 
The use of the nuclear spins surrounding electron spin qubits as  quantum registers 
and long-lived memories opens the way to new applications in quantum information and biological sensing.
Hence, there is a  need for generic and robust forms of control of the nuclear registers. Although adiabatic gates are widely used in quantum information, they can become too slow to outpace decoherence. Here, we introduce a technique whereby adiabatic gates arise from the dynamical decoupling protocols that simultaneously extend coherence. We illustrate this pulse-based adiabatic control for nuclear spins around NV centers in diamond. We obtain a closed-form expression from Landau-Zener theory and show that it reliably describes the dynamics. By identifying robust Floquet states, we show that the technique enables polarisation, one-shot flips and state storage for nuclear spins. These results introduce a new control paradigm that combines dynamical decoupling with adiabatic evolution.
\end{abstract}

\maketitle

There is enormous interest in the development of quantum technologies based on spins in the solid state. Optically active defects, such as the Nitrogen vacancy (NV) center in diamond, offer well-isolated individual electronic spins with long coherence times, optical addressability  and operation from cryogenic to room-temperatures \cite{Degen2014,Jelezko2004,Childress2006, Maze2008a,Bala2008,sivac1,sivac2}. In combination with nearby coupled nuclear spins, these systems provide a multi-qubit platform with a wide range of applications including detection, imaging and atomic-scale characterisation of the spin samples \cite{Cai,Muller14,Zhao2011,Zhao2012a,Renbao2014}, quantum computation and quantum networks \cite{NV3, Neumann2010,Bernien2013}. 

Both sensing and quantum information applications typically rely, for control of the central electronic spin, on sequences of periodically repeated microwave pulses, known as dynamical decoupling (DD).
Although DD was initially employed to decouple the central electron spin qubit from the decohering effect of the surrounding spin bath \cite{BarGill}, it was soon identified that the same sequences could also be used to sense individual nuclear spins \cite{Kolkowitz2012,Tam2012}.  For the case of the NV center in diamond, the entanglement generated between the electron spin and the I = 1/2 $^{13}$C nuclear spins is being explored as a tool for control, with nuclear spins as multi-qubit registers and quantum memories \cite{Taminiau2014,Registers}. Furthermore, state storage in nuclear spins has been used to augment protocols for quantum sensing and nanoscale NMR \cite{Wrachtrup2017}.

\begin{figure}[ht!] 
\includegraphics[width=2.5in]{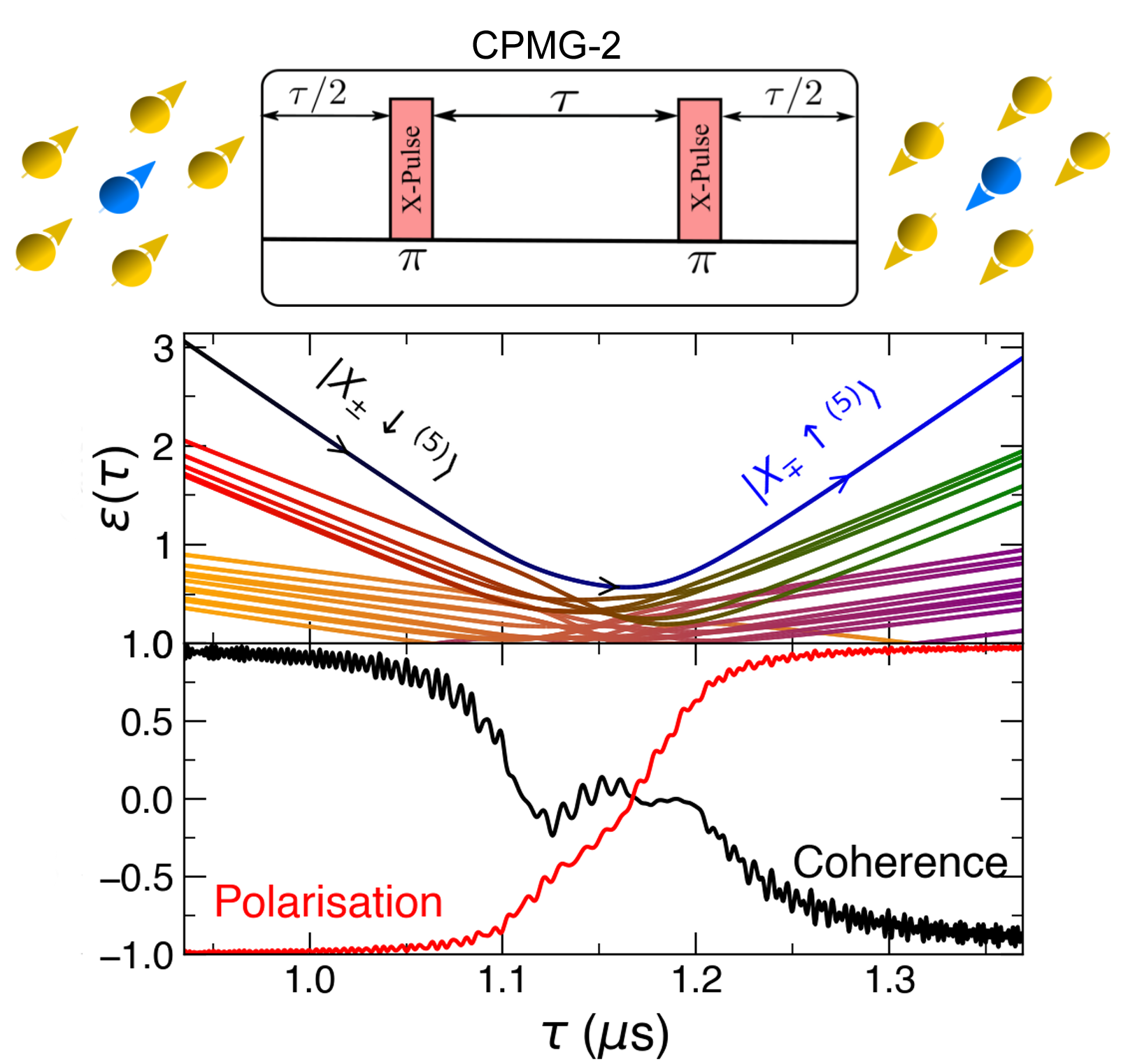}
\caption{Illustration of Ad-Pulse using the example of a NV electron spin system coupled to nearby nuclear spins: by means of  a slow adiabatic sweep over pulse interval $\tau$, one can coherently follow certain  robust eigenstates. The figure shows the eigenphases of Floquet states for the well-known CPMG-2N$_p$ dynamical decoupling (DD) protocol (illustrated above). The eigenphase spectra here is of a small cluster of 5 nuclear spins, $N_{nuc}=5$. For arbitrary $N_{nuc}$, the two fully polarised states  of the nuclei, $M_z=\frac{1}{\hbar}\sum_n \langle {\hat I}^{(n)}_z \rangle=\pm N_{nuc}/2$ where $n=1,2..N_{nuc}$, have a ``gap'' with the other $M_z$ manifolds. Thus a single  sweep over $\tau$ can flip an entire nuclear cluster from a $M_z=+N_{nuc}/2$ state to a $M_z=-N_{nuc}/2$ state, without requiring prior knowledge of the characteristic frequencies.  The NV electron spin 
retains coherence, in the above example evolving between the  $|X^\pm\rangle$ states at the endpoints.} 
\label{Fig1}
 \end{figure}
 
 \begin{figure*}[ht!]
\centering
\includegraphics[width=7in]{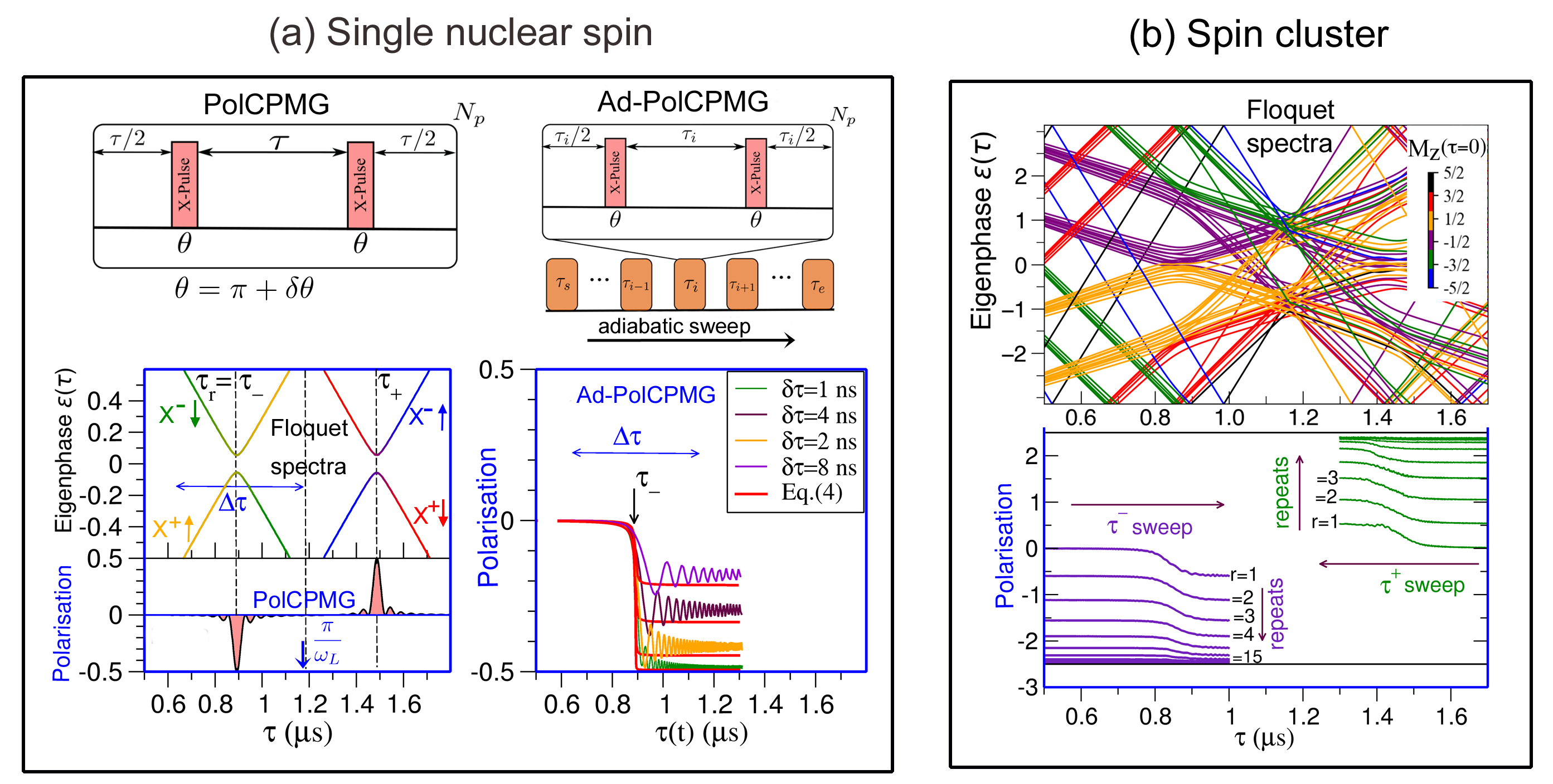}
\caption{ Comparison of standard DD with Ad-Pulse DD, using the PolCPMG polarisation protocol, applied to an NV centre electron coupled to one or more nuclear spins. In previously demonstrated PolCPMG experiments \cite{PolCPMG},  a slight under/overrotation of the $\pi$ pulses was shown to split the normal CPMG resonance at $\tau_r$ into two $\tau_r\equiv \tau^\pm$ resonances that are nuclear state selective.
The initial state is $|X^{\pm} \rangle_{NV} \otimes \rho_n $, so the nuclei are initially in a mixed state and $|X^{\pm}\rangle \equiv \frac{1}{\sqrt{2}}[|0\rangle \pm|1\rangle]$.
{\bf (a)  Single spin case.} DD resonances correspond to avoided crossings of the Floquet eigenphases (upper panel). 
Applying PolCPMG at $\tau^-$ or $\tau^+$ polarises the mixed state as shown (lower panel).
The  corresponding adiabatic protocol (Ad-PolCPMG) 
  sweeps slowly over a $\tau$ range of  $\equiv \tau^\pm\pm \Delta\tau/2$- without reinitialisation of the NV for different $\tau$. As the sweep step size, $\delta \tau$,  is reduced the sweep becomes adiabatic and full  polarisation is achieved: the behavior for all parameters is well described by 
 Eq.\ref{LZ}, based on Landau Zener theory (red curves).
{\bf (b) Multi-spin cluster}. The Floquet level structure is illustrated for the case of $N_{nuc}=5$
  nuclei. Although $M_z$ is not a good quantum number during the Ad-Pulse, for clarity, states are coloured according to the asymptotic $M_z$  at $\tau=0$. Polarisation can be  achieved even by sub-adiabatic sweeps $\Gamma_0 <1$ (see Eq.\ref{LZ})
   combined with repetitions (lower panel). Consequently, for each 
  repeat, the polarisation of the bath increases monotonically from zero for the initial nuclear mixture 
  to a fully polarised state for $\tau^-$. For $\tau^+$,  the sweep saturates, possibly because the end points are not asymptotic. Similar results with $\delta\tau= 1$ ns are obtained for any subset of the $N_{nuc}=7$ spin cluster investigated here. Coupling strengths are listed in \cite{Appendix}.}
 \label{Fig2}
\end{figure*}

Adiabatic quantum gates are well established in quantum information but, as adiabaticity entails  slow parameter sweeps,  outpacing decoherence represents a problem \cite{Awschalom}. Proposals employing  dynamical decoupling to extend
coherence {\em alongside} the adiabatic gates have been investigated \cite{Lidar,Witzel}. 
However, to our knowledge, the adiabatic gates that are  intrinsic to  the commonly used pulse-DD protocols
have not previously been considered: this may be surprising as they offer the important advantage that the coherence-protecting repeated pulse-DD protocols, in some sense, ``come for free''. 

Here we introduce and investigate a new technique that combines adiabatic passage with coherence protection through dynamical decoupling. We name this method Ad-Pulse.
This involves  adiabatically sweeping over the Floquet eigenphases of the DD pulse protocol. We show that this type of control is  quite generic and, for various applications, does not require knowledge of the individual resonances. Ad-Pulse can also be applied to a many-spin bath and we show some applications are insensitive to the number of spins.  We illustrate Ad-Pulse using the example of an NV centre and surrounding nuclear registers. We show also it may polarise and initialise small nuclear clusters  more effectively than the recently proposed  DD-based polarisation protocols PulsePol \cite{PulsePol,PulsePol1} or PolCPMG  \cite{PolCPMG} and we investigate  possibilities for  quantum state storage and read-out.

 Floquet theory is well-established in many fields in physics, mostly relating to continuous driving, including applications in  NMR \cite{FloquetNMR}.  However,  the term covers a wide range of scenarios:  the Floquet 
 theorem is applicable to any temporally periodic system.  One important application is so-called Floquet engineering (FE), \cite{Shirley1965} where a system driven by a typically strong or high frequency (non-resonant) field can be shown to correspond to an effective,  static Hamiltonian with renormalised parameters, by averaging over the period of the driving. Varying the {\em amplitude} of the non-resonant drive, one may tune over  the effective Hamiltonian. This approach has been proposed theoretically for polarisation of a nuclear spin bath \cite{PolChicago} using an adiabatic sweep of the energy eigenvalues of an effective static Hamiltonian.
 
The Ad-Pulse proposed here is quite distinct from FE:  there is no period averaging and in fact one sweeps
 coherently over the {\em period}  $T$ of the pulse protocol, over a range containing the characteristic resonances of the nuclear spins of interest. Previously it has been shown that the loss of coherence  identified in DD studies at resonances correspond to avoided crossings of the underlying Floquet eigenphases \cite{FloSpec,Flo2}.
Recognising this, Ad-Pulse  sweeps over these eigenphases: if sufficiently slow, the sweep follows each eigenstate  adiabatically and coherently.

For a system with a temporally periodic Hamiltonian,  ${\hat H}(t+T)={\hat H}(t)$ , Floquet's theorem allows one to write solutions of  the Schr\"odinger equation 
in terms  of quasi-energy states  $|\psi_l (t)\rangle= \exp{(-i{\epsilon}_l t)}   |\Phi_{l}\rangle$
where ${\epsilon}_l$ is the quasi energy,
$|\Phi_l(t) \rangle=|\Phi_l(t+T)\rangle$, $T$ is the period and
 $l=1,..,D$ ($D$ is the dimension of the state space).  

Where we require only ``stroboscopic'' knowledge of our system  at times $t= t+kT$ , one may obtain eigenstates of the one-period unitary evolution operator ${\hat U}(T) \equiv {\hat U}(T,0)$ for the joint electron-nuclear spin bath system under pulse-DD.
The Floquet states $ |\Phi_l\rangle$, obey the eigenvalue equation:
\begin{equation}
{\hat U}(T)|\Phi_l\rangle = \lambda_l  |\Phi_l\rangle \equiv  \exp{(-i{\mathcal E}_l)}   |\Phi_l\rangle
\label{Floquet}
\end{equation}
 where  ${\mathcal E}_l \equiv \tan^{-1}{ \operatorname{Im}{\lambda_l}/\operatorname{Re}{\lambda_l} }$ is the  eigenphase (the Floquet phase). 
 
 An NV electron spin system surrounded by $N_{nuc}$ nuclear spins is described by the Hamiltonian:
\begin{equation}
{\hat H}(t) = {\hat H}_p(t) +  \omega_\text{L} \sum_n {\hat I}^{(n)}_z + {\hat S}_z\sum_n\textbf{A}^{(n)} \cdot  {\bf I}^{(n)} 
\label{Ham}
\end{equation}

${\hat H}_p(t) = \Omega(t){\hat S}_x$ is the pulse control Hamiltonian. $\Omega(t)$ is the microwave drive strength which is non-zero during the pulses. For the CPMG sequence, the microwave pulses are applied along the $x$-axis at regular intervals, $\tau$, as shown in Fig.~\ref{Fig1}. The pulse duration for a $\pi$-flip of the electron spin  is denoted $T_\pi$. $\omega_\text{L}$ is the nuclear Larmor frequency; the hyperfine field  $\textbf{A}^{(n)}$ felt by the nuclear spin has components
 $A^{(n)}_\perp, A^{(n)}_z$ relative to the $z$-axis. Without loss of generality, we take ${A^{(n)}_\perp} \equiv A^{(n)}_x$.
 We omit the $n$ superscript for single-nuclear spin calculations below and denote as $\tau_r$ 
the resonant pulse spacing for which the electron and nuclear spins entangle.
For CPMG, $\tau_r =j \pi/(\omega_L+A_z/2)$, where $j$ is an odd integer.
 
 In Fig.\ref{Fig1} we illustrate an application of Ad-Pulse using the well-known CPMG-2N$_p$  dynamical decoupling protocol.  We consider a bath of 5 nuclear spins, C1,C2...C5 (for list of couplings see table in \cite{Appendix}) from a cluster employed as registers 
 in recent experiments in \cite{Registers},  with $A_x/(2\pi)$ within the range $[20:60]$ kHz.
  The Floquet eigenphases ${\mathcal E}_l \equiv {\mathcal E}_l (\tau)$ are plotted over a
 range centered around $\tau = \pi/\omega_L$, for $B_0=0.0403$ T. 
We plot  ${\mathcal E}_l \in [-\pi:\pi]$ noting the spectra are multiply degenerate under the shift ${\mathcal E}_l \to {\mathcal E}_l + 2\pi n$, for integer $n$ \cite{Flo2}.

 In a standard DD-based sensing protocol, $N_p$ cycles would be applied at each value of $\tau$. The NV electronic spin is reinitialised optically after each value of $\tau$. Ad-Pulse instead sweeps over a range $\Delta \tau$ of pulse spacings without reinitialisation, moving adiabatically on a single Floquet eigenstate in $k=1,...N_{s}$  steps. Thus $\tau_{k+1}= \tau_k+ \delta \tau$ where $\delta \tau$ is the step in $\tau_k$ and the total sequence time up to step $k$ is:
 \begin{equation}
t_k= \sum_{l=1}^{l=k } 2\tau_l N_p  \  \textrm{where }  \  \tau_k \equiv \tau(t_k). 
\label{Sweep}
\end{equation}
For a coherent sweep, the total sweep time is $t_{Tot}=\sum_{l=1}^{l=N_{s} } 2\tau_l  N_p$ 
and is required to be less than the NV centre's coherence time under a DD protocol, $T_2$. 

A generic feature of the  Floquet spectra, for arbitrary number of of nuclei $N_{nuc}$,  is the gap seen
between two extremal states which asymptotically tend to maximal/minimal polarised states 
$M_z = \pm N_{nuc}/2$; and the other $M_z$ manifolds.  Each and
every one of the states in the adjoining $M_z=\pm(N_{nuc}/2-1)$ manifolds
experiences a coupling $A_x^{(n)}  {\hat I}_x^{(n)} {\hat S}_z $  associated with an anti-crossing \cite{Flo2} 
- and level repulsion- with the extremal states.  Hence there is {\em always } a gap with  the adjoining manifolds.

The lower panel of Fig.\ref{Fig1} shows the NV spin coherence $\mathcal{L}(\tau)=\langle {\hat S}_x\rangle$ and 
nuclear bath polarisation $\mathcal{P}(\tau)=\frac{1}{N_{nuc}}\sum_n \frac{2}{\hbar} \langle {\hat I}^{n}_z\rangle$ as an adiabatic sweep over $\tau$ is carried out, using the common CPMG protocol. The sweep over a single eigenstate  is shown to invert the entire nuclear bath : an initial  ``down'' polarised bath state $|X_\pm\downarrow^{(N_{nuc})}\rangle$ follows the Floquet state trajectory to a fully `up'  polarised bath state $|X_\mp\uparrow^{(N_{nuc})}\rangle$. Provided  $t_{Tot} \lesssim T_2$, a coherent NV state evolves into another coherent NV state (from $|X_+\rangle $ to $|X_-\rangle $). 
Unlike  the FE polarisation method \cite{PolChicago}, the total bath polarisation $M_z\hbar= \sum_n \langle I^{(n)}_z \rangle$ is not a good quantum number, allowing the whole-bath flip illustrated in  Fig.\ref{Fig1}.

We can use Landau-Zener theory to  analyse the adiabatic sweep  from an initial $\tau=\tau_{ini}$ to 
a final $\tau_{fin}$.  The range $\tau_{ini}:\tau_{fin}$ contains the anticrossing region near $\tau\simeq \tau_r$, but $\tau_{ini},\tau_{fin}$  both lie {\em outside} it. For a single nuclear spin,   for any value of $\tau \in [\tau_{ini}:\tau_{fin}]$, we find the probability of losses from the eigenstate trajectory are given by $ e^ {-\Gamma_{LZ}(\tau(t))} $, where:
 \begin{eqnarray}
\Gamma_{LZ}(\tau(t)) &=&  \frac{2 A^2_x \tau_r^2T_r}{\beta^2 \delta \tau}  F(\tau_{ini},\tau)  \equiv  \Gamma_0 F(\tau_{ini},\tau) \ \textrm{where} \nonumber \\
 F(\tau_{ini},\tau) &=&  \frac{1}{\pi} \left[\arctan{(\Phi_\tau)} - \arctan{(\Phi_{\tau_{ini}})}\right]
\label{LZ}
\end{eqnarray}
and $\Gamma_0$ is the well-known Landau Zener exponent (see  \cite{Appendix} for further details).
 $\Phi_\tau=\frac{4\pi(\tau-\tau_r)}{\tau} \sqrt{T/\delta\tau}$. As $\tau \to \tau_{fin}$, then  $F(\tau_{ini},\tau) \to 1$ 
and $\Gamma_{LZ}(\tau(t)) \to \Gamma_0$.
We can readily relate Eq.\ref{LZ}  to a corresponding nuclear polarisation using $P_{pol}(t)\simeq \pm(1-e^{-\Gamma_{LZ}(t)})/2 $  that here we can give in closed form in terms of the couplings $A_z,A_x$, sweep parameters, applied magnetic field
and may compare with exact numerics. 

For adiabaticity (no losses)  $\Gamma_0 >>1$ and this requirement constrains $t_{Tot}$. However, the sweep range $\Delta \tau=\tau_{Ns}-\tau_1$ required is another important consideration: i.e., although $\Gamma_0 \propto A_x^2$, since  $\Delta \tau \sim A_x$, we find the total time required increases linearly  $t_{Tot} \propto  A_x^{-1}$ \cite{Appendix}. For optimised Ad-Pulse, from Eq.\ref{Sweep}, we  can take $\tau_1\equiv \tau_{ini}$ and $\tau_{fin}\equiv \tau_{N_s}$. 

$\beta$ is a protocol dependent parameter.  We investigated adiabatic control in CPMG as well as the DD polarisation protocols PulsePol and PolCPMG.  PulsePol is a recently proposed robust nuclear polarisation technique with a multi-pulse protocol \cite{PulsePol,PulsePol1}. For pulse spacings $\tau_r \simeq j \pi/(2\omega_I)$, where $j=1,3...$ it yields an  effective  flip-flop Hamiltonian ${\hat H}_{PP} = g I^\pm S^\mp$.   PolCPMG  on the other hand, is a variant of CPMG,  obtained by applying an over/under-rotation to the pulses i.e. $\theta=(1+\delta\theta) \pi$ where $\delta \theta=0$ corresponds to CPMG.  It was also shown experimentally to hyperpolarise a nuclear bath \cite{PolCPMG} for $\delta \theta \simeq 0.05-0.25\pi$.  $\beta=  (\pi+\delta\theta)$ for Ad-PolCPMG/CPMG  while $\beta=6\pi/(2+\sqrt{2})$ for Ad-PulsePol.  $T_r = 4\tau_r$ for PulsePol, $T_r = 2\tau_r$ for CPMG/ PolCPMG.

Fig.\ref{Fig1} assumed nuclei  start from a pure,  polarised initial state. Generally the nuclear bath is in a mixture; initialising the NV spin can be achieved optically and is comparatively straightforward.
In  Fig.\ref{Fig2} we show  how Ad-Pulse adiabatic sweeps with PolCPMG or PulsePol may also be used to prepare a pure initial nuclear state from a mixture. We show results for  Ad-PolCPMG, but Ad-PulsePol yields similar behaviours.

Fig.\ref{Fig2}a (left panels) shows  the Floquet spectrum for PolCPMG  for a single spin for $\delta \theta= 0.25\pi$. The degeneracy of the level anticrossings in CPMG  is lifted into two distinct  crossings at $\tau_\pm$, and hence there are two coherence dips in the NV's trace. Applying PolCPMG at $\tau_-$ ($\tau_+$) will target the $|\uparrow\rangle$ ($|\downarrow\rangle$) nuclear state and evolve it into the $|\downarrow\rangle$ ($|\uparrow\rangle$) state, polarising the spin. Fig.\ref{Fig2}(a) (right panels) shows the application of Ad-PolCPMG to initialise a single spin and shows the polarisation is well-described by Eq.\ref{LZ}: full polarisation is achieved when the sweep is adiabatic. In fact, if we allow repeated sweeps/reinitialisation of the NV electron spin, we find the spin bath polarisation is  robust to a degree of non-adiabaticity. 

Fig.\ref{Fig2}(b) examines the case of multiple nuclear spins.  
The right panel shows the multi-spin Floquet spectra. The levels group into manifolds of different $M_z$ (for the  5 spins case,  $M_z= -5/2,-3/2...+5/2$ and there are 6 manifolds). Hyperfine coupling only allows couplings (or avoided crossings) between neighbouring $M_z$ and $M_z \pm 1$ manifolds.

For the multi-spin Ad-PolCPMG, and the  $\tau=\tau^-$ resonance, we find that about $r=10-15$  repetitions is sufficient to fully polarise $(>99.5 \%)$, whereas standard PolCPMG requires 100s of repeated sequences and NV re-initialisation. For $\tau^+$, a limit corresponding to about $95\%$ polarisation is reached, possibly because the initial and end states do not correspond
to the asymptotic $|X^\pm\rangle |M_z\rangle$ basis. A sweep from $\tau=0$ avoids these issues.
 Similar behaviour was observed for clusters with  $N_{nuc}$ up to 7. 
 
 If the sweeps are far from adiabatic
 we may combine the contributions from Eq.(\ref{LZ}) additively $P_{pol}(t)= \sum_n P^{(n)}_{pol}(t)$ to accurately describe full simulations. The converse is not true: the highly adiabatic $\delta \tau =1$ ns sweeps in Fig.(\ref{Fig2})(b) are not described by Eq.\ref{LZ}: each sweep achieves at most the equivalent of 1 spin-flip (as may be seen from the equispaced $r=1,2,3$ sweeps). This is in contrast to
 the pure state trajectory in  Fig\ref{Fig1} where for the extremal state, the $X^+ \to X^-$ corresponds to a whole bath flip.

%Unlike standard DD methods (PolCPMG or PulsePol) , polarisation using this method requires no knowledge 
%of the resonant frequencies $\pi/\tau^{(n)}_r$ of the $^{13}$C nuclear spins. 

\begin{figure}[t!]
\includegraphics[width=3.in]{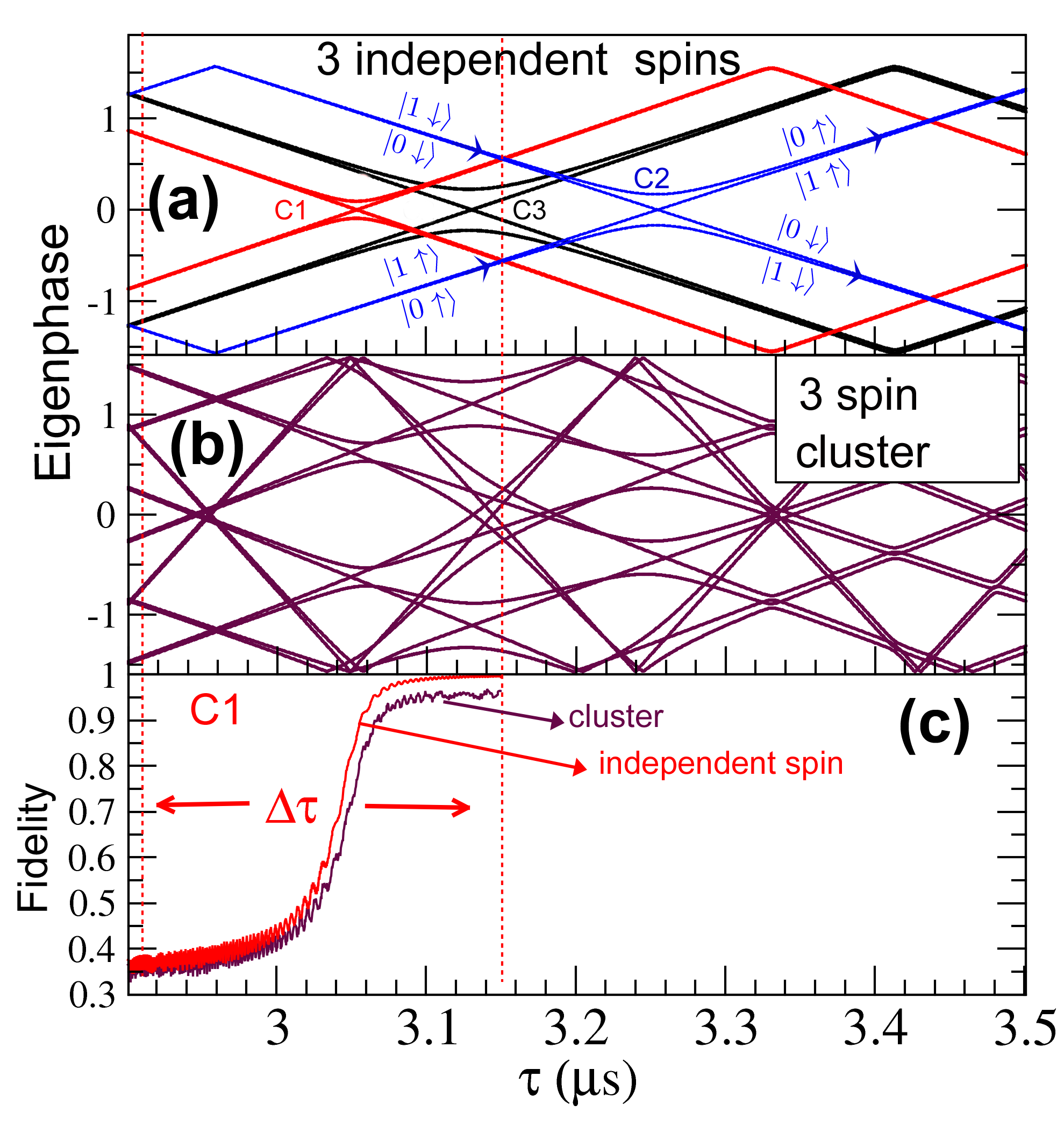}
\caption{ Illustration of state storage and read-out using the Ad-PulsePol protocol. {\bf(a)} The top panel
shows the Floquet spectra for the three independent spins C1-C3. For this protocol it is insightful to show 
eigenphases ${\mathcal E}_l \in [-\pi/2:\pi/2]$ as the spectra show three pairs of crossings: each of these corresponds to a true crossing superposed on an avoided crossing. {\bf(b)} Spectra for the full 3-nuclear spin dynamics.
Each pair of crossings in (a) is now associated with a family of  4 anti-crossings.
 A single Ad-sweep over one of crossings (we investigate the $\tau=0.7-1.0 \  \mu$s
range), transfers an arbitrary NV state $|\psi \rangle_{NV} \otimes |\downarrow/\uparrow\rangle $ 
to a pure nuclear state,  $|0/1 \rangle_{NV} \otimes |\psi\rangle_n $, within Larmor phases.   {\bf(c)}  Shows the fidelity of the state transfer for  C1 ($A_x=26.6 \times 2\pi $ kHz). }
% In the latter case a nucleus with the lowest value of $\tau$ for
%the resonance was chosen. }
%The technique has high fidelity for clusters with very few strong-coupled nuclei, but
%is not robust for generic larger clusters.}
\label{Fig3}
\end{figure}

In Fig.\ref{Fig3}  we investigate Ad-PulsePol for quantum state storage. The corresponding Floquet spectrum is shown in Fig.\ref{Fig3} for the three nuclei C1-C3, either as independent spins (top panel) or for the  3-spin cluster (lower panel), for $B_0=0.0403$ T.  For Ad-PulsePol, plotting the eigenphases ${\mathcal E}_l \in [-\pi/2:\pi/2]$ is most insightful,  as it shows that the important crossings come in pairs. They are labelled for spin C2 (blue) 
for the third harmonic ($j=3$ for $\tau_r$ ) corresponding to ${\hat H}_{PP} = g I^+ S^-$. They
 show that an initial state $|1 \downarrow\rangle \leftrightarrow |0 \uparrow\rangle$,  while 
 $|1 \uparrow\rangle  \rightarrow |1 \uparrow\rangle$
and $|0 \downarrow\rangle \to |0 \downarrow\rangle$. 

Finally, we show that an adiabatic sweep passing through such a pair of crossings would map an arbitrary initial NV state onto the nuclear spin state:
 \begin{eqnarray}
 &[a|0\rangle+b|1\rangle]_{NV}\otimes & |\downarrow\rangle_{nuc} \to |0\rangle_{NV} \otimes  [a|\downarrow\rangle+b|\uparrow\rangle]_{nuc} \nonumber \\
& [a|0\rangle+b|1\rangle]_{NV}\otimes & |\uparrow\rangle_{nuc} \to  |0\rangle_{NV} \otimes  [a|\downarrow\rangle+b|\uparrow\rangle]_{nuc}\nonumber \\
\label{Stor}
\end{eqnarray}
noting that for the second line, the adiabatic sweep leaves the NV in the $|1\rangle$ state, 
so optical reinitialisation to $|0\rangle$ is assumed. A $Z$ gate on the nuclei is also assumed: the relative nuclear phases rotate as $e^{\pm i \omega_L t}$ thus may be chosen by timing the Larmor precession. 

We test this for an adiabatic sweep over $\Delta \tau$ centered on $\tau_{PP}$ for spin C1 ($A_x/(2\pi) \simeq 20$ kHz), starting with the test state 
 $|\Psi_0\rangle= [\frac{1}{\sqrt{3}}|0\rangle+\frac{2}{\sqrt{3}}|1\rangle]_{NV}\otimes |\downarrow\rangle_{nuc}$ and testing
 the fidelity of the overlap $\mathcal{F}(\tau) =|\langle \Psi_0|\Psi_T\rangle|$ where   \\$|\Psi_T\rangle=|0\rangle_{NV}\otimes[\frac{1}{\sqrt{3}}|\downarrow\rangle+\frac{2}{\sqrt{3}}|\uparrow\rangle]_{nuc}$, but disregarding phases between the nuclear states for convenience (by taking modulus of the coefficients of $\Psi_T$).
  The partial overlap between avoided crossings reduces the fidelity appreciably, indicating that although this technique may be used, it is restricted to very well isolated (well-resolved in $\tau_{PP}$) nuclear spins. 
Ad-PolCPMG can equally achieve an equivalent state storage gate. 
 
{\bf Conclusions} 
We introduce and discuss the new technique of adiabatic dynamical decoupling and have shown it has a broad range of potential applications that opens several avenues for experimental studies, ranging from polarisation and initialisation of nuclear mixtures to control of single multi-spin pure states for state storage and preparation. There are many different DD protocols and further  studies will be required to identify new types of robust states, including engineered many body Hamiltonian approaches \cite{Manybody}, that will allow adiabatic coherent control on multiple bath spins simultaneously.

{\bf Acknowledgements} 
The authors thank Conor Bradley and Asier Galicia for useful discussions. Oliver Whaites acknowledges support from an EPSRC DTP studentship grant. This work was supported by the Netherlands Organisation for Scientific Research (NWO/OCW) through a Vidi grant and as part of the Frontiers of Nanoscience (NanoFront) programme and the Quantum Software Consortium programme (Project No. 024.003.037/3368). This project has received funding from the European Research Council (ERC) under the European Union’s Horizon 2020 research and innovation programme (grant agreement No. 852410).

\end{document}